\documentclass[a4paper,english,useAMS,usenatbib]{mn2e}
\usepackage[T1]{fontenc}
\usepackage[latin9]{inputenc}
\usepackage[active]{srcltx}
\usepackage{color}
\usepackage{prettyref}
\usepackage{amsmath}
\usepackage{amssymb}
\usepackage{graphicx}
\usepackage{esint}
\usepackage[authoryear]{natbib}

\makeatletter

\providecommand{\tabularnewline}{\\}

\@ifundefined{date}{}{\date{}}

\newcommand{\nat}{Nature}
\newcommand{\mnras}{Monthly Notices of the Royal Astronomical Society}
\newcommand{\apj}{The Astrophysical Journal}
\newcommand{\apjl}{The Astrophysical Journal Letters}
\newcommand{\apjs}{The Astrophysical Journal Supplement Series}
\newcommand{\aap}{Astronomy and Astrophysics}
\newcommand{\aj}{The Astronomical Journal}
\newcommand{\araa}{Annual Review of Astronomy and Astrophysics}
\newcommand{\pasa}{Publications of the Astronomical Society of Australia}
\newcommand{\hi}{\textsc{H\,i}}

\title[{\hi} in distant galaxies using spectral stacking]{\textbf{Detection of {\hi} in distant galaxies using spectral stacking}}\author[J. Delhaize et al.]{J. Delhaize$^{1}$\thanks{E-mail:jacinta.delhaize@icrar.org}, M. J. Meyer$^{1,2}$, L. Staveley-Smith$^{1,2}$, B. J. Boyle$^{2,3}$\\$^{1}$International Centre for Radio Astronomy Research (ICRAR), M468, University of Western Australia, 35 Stirling Hwy, Crawley, WA, 6009, Australia\\$^{2}$ARC Centre of Excellence for All-sky Astrophysics (CAASTRO)\\$^{3}$CSIRO Astronomy and Space Science, PO Box 76, Epping, NSW 1710, Australia}

\makeatother

\usepackage{babel}
\begin{document}

\date{Accepted 2013 May 8. Received 2013 May 6; in original form 2013 February
23}

\maketitle
\pagerange{\pageref{firstpage}--\pageref{lastpage}} \pubyear{2013}

\label{firstpage}
\begin{abstract}
Using the Parkes radio telescope, we study the 21\,cm neutral hydrogen
({\hi}) properties of a sample of galaxies with redshifts $z<0.13$
extracted from the optical Two-Degree-Field Galaxy Redshift Survey
(2dFGRS). Galaxies at $0.04<z<0.13$ are studied using new Parkes
observations of a 42\,deg$^{2}$ field near the South Galactic Pole
(SGP). A spectral stacking analysis of the 3,277 2dFGRS objects within
this field results in a convincing 12\,$\sigma$ detection. For the
low-redshift sample at $0<z<0.04$, we use the 15,093 2dFGRS galaxies
observed by the {\hi} Parkes All-Sky Survey (HIPASS) and find a 31\,$\sigma$
stacked detection. We measure average {\hi} masses of $(6.93\pm0.17)\times10^{9}$
and $(1.48\pm0.03)\times10^{9}$\,$h^{-2}\,{\rm M}_{\odot}$ for
the SGP and HIPASS samples, respectively. Accounting for source confusion
and sample bias, we find a cosmic {\hi} mass density of $\Omega_{\hi}=(3.19_{-0.59}^{+0.43})\times10^{-4}\, h^{-1}$
for the SGP sample and $(2.82{}_{-0.59}^{+0.30})\times10^{-4}\, h^{-1}$
for the HIPASS sample. This suggests no (12$\pm$23 per cent) evolution
in the cosmic {\hi} density over the last $\sim$1\,$h^{-1}$\,Gyr.
Due to the very large effective volumes, cosmic variance in our determination
of $\Omega_{\hi}$ is considerably lower than previous estimates.
Our stacking analysis reproduces and quantifies the expected trends
in the {\hi} mass and mass-to-light ratio of galaxies with redshift,
luminosity and colour.
\end{abstract}

\section{Introduction}

To obtain a complete picture of galaxy evolution, it is crucial to
understand how the cold gas content of galaxies varies with cosmic
time. Cold gas ($<10^{4}$\,K), largely in the form of neutral hydrogen
({\hi}), is supplied through various accretion, merger and feedback
processes (eg. \citealt{keres05}) and is later available to condense
into massive molecular clouds and stars. Without this gas, star formation
is quenched and evolution continues only passively. To improve our
understanding of the evolution in quantities such as star formation
rate density (\citealt{madau96}, \citealt{hopkins06}), it is therefore
fundamental to also improve our understanding of changes in the cosmic
gas density. 

Models that attempt to predict the variation of cosmic gas density
with redshift diverge in their predictions. This is partly due to
the wide spatial range required to track the requisite gas physics
and partly because of uncertainty in the physics of galaxy formation,
in particular the action of feedback from stars and AGN (eg. \citealt{somerville01,cen03,nagamine05,power10,lagos11}).
Thus, observational constraints are vital.

A number of observational techniques exist for tracing {\hi} gas
within galaxies. At high redshifts, the Gunn-Peterson effect applied
to damped Lyman $\alpha$ (DLA) systems can be exploited. \citet{prochaska05}
and \citet{prochaska09} used Sloan Digital Sky Survey (SDSS) spectra
to trace systems with {\hi} column densities above $2\times10^{20}$\,cm$^{-2}$
at $z>1.7$. Though they find a statistically significant detection
of evolution in the cosmic {\hi} density within $2.2<z<3.5$, interpretation
of their results is problematic due to systematic effects such as
dust obscuration and gravitational lensing. At $z<1.65$ the Lyman
$\alpha$ transition is observed at ultraviolet wavelengths, making
it difficult to trace with the limited time available on space-based
observatories. \citet{rao06} were able to contribute in this regime
by optically identifying DLA systems through metal absorption lines
in SDSS data. They found no evidence of evolution in the cosmic {\hi}
density of DLAs within $0.5<z<5$, although their results are limited
by statistical uncertainty.

The preferred method for tracing {\hi} in the local Universe ($z\approx0$)
is via the direct detection of the neutral hydrogen 21\,cm hyperfine
emission line. With the wide instantaneous field of view provided
by multibeam receivers on large radio telescopes, it is feasible to
conduct blind, all-sky surveys at 21\,cm. This has facilitated the
detection of {\hi} in large numbers of galaxies. A prime example
of this is the {\hi} Parkes All-Sky Survey (HIPASS), conducted with
the 64\,m Parkes radio telescope in New South Wales, Australia \citep{barnes01,staveley-smith96}.
HIPASS detected {\hi} within $5,317$ galaxies at $0<z<0.04$ \citep{meyer04,wong06},
allowing the construction of the most reliable and complete {\hi}
mass function then available, and an estimate of the {\hi} mass density
for the local Universe \citep{zwaan05}. Similarly, the Arecibo Legacy
Fast ALFA (ALFALFA) survey aims to detect >30,000 galaxies at 21\,cm
out to $z=0.06$ over a sky area of $\sim$7,000 deg$^{2}$ \citep{giovanelli05}.
Using the 40 per cent ALFALFA catalogue, \citet{martin10} find a
cosmic {\hi} mass density 16 per cent higher than that derived by
\citet{zwaan05}. They conclude that this discrepancy is caused by
an under-representation in the HIPASS catalogue of the rare, high-mass
{[}$9.0<\log(M_{{\rm {HI}}}/M_{\odot})<10.0${]} galaxies due to sensitivity
limits.

Deeper, blind {\hi} surveys are underway, such as the Arecibo Ultra
Deep Survey (AUDS). This survey aims to directly detect significant
numbers of field galaxies out to $z=0.16$ to probe the evolution
of cosmic {\hi} properties \citep{freudling11}. However, the survey
time required to achieve sufficient sensitivity at these higher redshifts
restricts the observations to a narrow field (1/3\ deg$^{2}$ for
AUDS), thus potentially introducing a strong cosmic variance bias.

To date, the deepest detections of 21\,cm emission in individual
galaxies have been made at $z\approx0.2$ by \citet{catinella08}
using pointed observations with the sensitive 305\,m Arecibo telescope,
and by \citet{zwaan01} and \citet{verheijen07} using long integrations
of cluster galaxies with the Westerbork Synthesis Radio Telescope
(WSRT). While future instruments such as the Square Kilometre Array
(SKA) and its pathfinders should have the ability to detect 21\,cm
emission to much higher redshifts over large sky areas, these will
not be available for a number of years.

However, methods other than direct, individual detections are available
to push the current redshift limit of {\hi} observations. In particular,
stacking is a useful method for probing {\hi} within large samples
of distant galaxies. Stacking is the process of co-adding individual
non-detections in an attempt to boost the signal-to-noise ratio (S/N)
of the data and thereby recover a more significant statistical detection. 

\citet{zwaan00}, \citet{chengalur01} and \citet{lah09} demonstrated
that stacking can be successfully used to examine the environmental
influence on the {\hi} properties of cluster galaxies out to $z=0.37$.
\citet{lah07} were the first to apply this co-adding method to field
galaxies to study the evolution of the cosmic {\hi} density. They
examined 154 star-forming galaxies selected in H$\alpha$ at $z=0.24$
using 21\,cm data from the Giant Metrewave Radio Telescope (GMRT);
however, averaged detections were 3\,$\sigma$ at best. 

\citet{fabello11} and \citet{fabello11b} used similar stacking methods
to examine the impact of bulge presence and AGN activity on the {\hi}
content of galaxies with stellar masses greater than 10$^{10}$\,$M$$_{\odot}$.
They demonstrated that when using relatively low redshift ($0.025<z<0.05$)
{\hi} data and a large sample of galaxies, strong stacked detections
were possible, even for the subsample of galaxies individually undetected
at 21\,cm.

The related technique of `intensity mapping', which uses cross-correlation
of radio and optical redshift data, has been used by \citet{pen09},
\citet{chang10} and \citet{masui13} to detect {\hi} in samples
at even higher redshift.

The work presented here aims to constrain the cosmic density of {\hi}
out to $z=0.13$ and demonstrate the accuracy of the stacking method
over this redshift range. We have employed the fast survey speed of
the Parkes telescope to collect 21\,cm observations of a very large
sample of field galaxies from the Two-Degree-Field Galaxy Redshift
Survey (2dFGRS). The sample selection is not biased by environment,
star formation signatures, or any particular physical characteristic
other than the optical magnitude limits of the 2dFGRS. By using the
wide but low-redshift HIPASS data, in combination with new, spectrally
deep Parkes observations of a smaller subsection of the 2dFGRS footprint,
we can examine the {\hi} content of galaxies over the full range
$0<z<0.13$.

Section 2 of this paper presents the optical redshift survey and the
radio observations used in the analysis. The source selection and
stacking procedures are described in Section 3. In Section \ref{sub:Noise-Behaviour}
we present an analysis of noise behaviour in the {\hi} data. Section
\ref{sub:Confusion} discusses the influence of source confusion on
our results and illustrates how we correct for this. Section \ref{sub:Cosmic-HI-mass}
describes the {\hi} mass density calculation and Section \ref{sub:Cosmic-variance}
considers the impact of cosmic variance. Trends in {\hi} properties
with redshift, luminosity and colour are measured in Section \ref{sub:Binning}.
Finally, we present our conclusions in Section 5. We have assumed
a $\Lambda$ cold dark matter cosmology with a reduced Hubble constant
$h=H_{0}/$(100\,km\,s$^{-1}$\,Mpc$^{-1}$)$=1.0$, $\Omega_{\Lambda}=0.7$
and $\Omega_{{\rm M}}=0.3$.

\section{The data\label{sec:The-data-1}}

\subsection{The 2dFGRS optical catalogue\label{sub:2dFGRS-optical-catalogue}}

The basic principle of the {\hi} stacking analysis employed here
is to identify the positions and redshifts of galaxies using an external
source catalogue containing optical redshifts, extract the {\hi}
information at these coordinates and then co-add the data. For the
present analysis, the positions and redshifts are provided by the
2dFGRS. The 2dFGRS is a spectroscopic survey of over 250,000 galaxies
conducted with the 2dF multifibre spectrograph on the Anglo-Australian
Telescope. This survey covers two large strips, one in the Northern
Galactic hemisphere and one in the south, as well as a number of small,
random fields. The total sky coverage is $\sim$2,000 deg$^{2}$.
See \citet{colless01} for further details of the surveyed regions.

The input photometric source catalogue for this survey comes from
an adapted catalogue of Automated Plate Measuring (APM) machine scans
of the Southern Sky Survey taken with the UK Schmidt Telescope. These
target sources have extinction-corrected magnitudes brighter than
$b_{J}=19.45$. All APM $b_{J}$ magnitudes used in this paper are
extinction-corrected, but not $(k+e)$-corrected for bandpass and
evolution. These corrections will be incorporated into the corresponding
redshifted luminosity density.

As described in \citet{colless01}, the redshifts were estimated using
a combination of automated absorption spectral shape fitting, emission
line identification and manual input. Each redshift was assigned a
quality factor $Q$ between 1 (no reliable redshift available) and
5 (very reliable redshift). We find that sources with $Q\geqslant3$
are sufficient for our stacking analysis. \citet{colless01} state
that $Q\geqslant3$ redshifts have a root-mean-square (rms) uncertainty
of 85\,km\,s$^{-1}$ and can be used with 98.4 per cent reliability.

\subsection{{\hi} observations and data\label{sub:HI-data}}

\subsubsection{South Galactic Pole observations\label{sub:pks_data}}

We conducted a 21\,cm survey with the 64\,m Parkes radio telescope
over a 42\,deg$^{2}$ ($7^{\circ}$$\times$$8^{\circ}$) sky field
centred on right ascension (RA) 00$^{h}$42$^{m}$00$^{s}$ and declination
(Dec.) -29$^{\circ}$00$'$00$''$ (J2000), close to the South Galactic
Pole (SGP). The region was chosen as it contains an overdensity of
available 2dFGRS spectra (described further in Section \ref{sub:Source-selection-sgp}).
This provides a high number of spectra per square degree available
for a stacking analysis. 87 h of telescope time was used for observations
of this field. These were conducted on 2008 November 20-25, 2009 August
17-20 and 2009 September 25-29 with an estimated 52 h of on-source
integration. 

The observations used the multibeam correlator with two adjacent frequency
bands, each with 1024 spectral channels over a bandwidth of 64\,MHz.
The bands were centred on 1285 and 1335\,MHz with a 14\,MHz overlap
to reduce the impact of the bandpass shape. Observations in the two
frequency bands were taken independently. The full frequency coverage
of the Parkes data is thus 1253-1367\,MHz over two polarizations.
We therefore have the potential to detect galaxies with 21\,cm redshifts
in the range $0.0391<z<0.1336$. The resulting channel spacing is
62.5\,kHz.

\subsubsection{HIPASS observations\label{sub:HIPASS-data}}

We also present a study using all overlapping data from the 2dFGRS
catalogue and HIPASS. HIPASS was also conducted with the Parkes 64\,m
dish using the 21\,cm multibeam receiver \citep{staveley-smith96}
and covers the entire sky south of Dec. +25$^{\circ}$. The HIPASS
velocity range is $-1280<cz<12700$\,km\,s$^{-1}$, corresponding
to a frequency coverage of 1362.5$-$1426.5\,MHz and redshifts below
$z=0.0423$. Observations of the southern field (below Dec. +2$ $$^{\circ}$)
were completed over the period from 1997 February to 2000 March \citep{barnes01}.
The northern field (Dec. +2$^{\circ}$ to +25$^{\circ}$) was observed
from 2000 to 2002 \citep{wong06}. The HIPASS and SGP data sets overlap
by only 4.5\,MHz ($\Delta z=0.003$) in spectral range, making the
two data sets highly complementary.

\subsubsection{Telescope scanning mode}

Both data sets were collected using a similar Parkes scanning mode,
where the telescope is actively scanned across the required sky area
at a rate of 1$^{\circ}$\,min$^{-1}$. While HIPASS only used scans
in declination, the SGP observations used a `basket-weave' pattern,
with the entire field also covered by right ascension scans. This
reduces the effect of negative sidelobes around bright continuum sources
created by the bandpass correction. Using this technique, the SGP
field was completely surveyed 15 times in each spectral band within
the available time frame.

\subsubsection{Data reduction and interference mitigation}

Bandpass removal and calibration were conducted identically for both
data sets using the \textsc{livedata}%
\footnote{\textsc{livedata} is supported by the Australia Telescope National
Facility and is available at http://www.atnf.csiro.au/computing/software/livedata/%
} reduction package. See \citet{barnes01} for a detailed explanation
of these processes. The data were smoothed to reduce the effects of
ringing created by strong Galactic emission and/or strong radio frequency
interference (RFI), such as from satellites. Hanning smoothing was
applied to the SGP data. HIPASS utilized Tukey 25 per cent smoothing.
However, we have also applied Hanning smoothing to the data for consistency.
The spectral resolution after smoothing is 125\,kHz (26.4\,km\,s$^{-1}$
at $z=0$).

\begin{figure}
\includegraphics[bb=20bp 10bp 530bp 400bp,clip,scale=0.45]{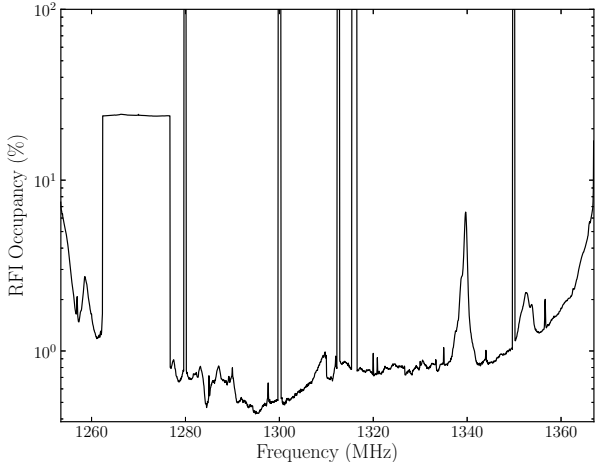}\caption{The percentage of low-quality SGP data excluded in each frequency
channel, largely due to the presence of satellite RFI.\label{fig:rfi-occupancy}}

\end{figure}

The use of robust statistics in HIPASS data processing (see Section
\ref{sub:Gridding} below) was sufficient to mitigate against the
majority of RFI. However, the lower frequencies probed during SGP
observations suffer from greater RFI. To improve the quality of the
SGP data, we applied an automated RFI masking routine to the reduced,
smoothed data. The mean and rms noise levels were calculated in an
RFI-free region and the resultant $3\,\sigma$ clipping threshold
was set at 250.5\,mJy. All pixels with absolute deviations above
this threshold were masked. This was not sufficient to completely
mitigate against the presence of strong satellite RFI in the frequency
range 1262.4$-$1276.8\,MHz. It was not consistently present and
appears in $24$ per cent of the observations in the 1285\,MHz band.
Therefore, the full range of affected channels was completely masked
in all data files containing this RFI. This method was successful
in recovering only the good quality data in the broad-band RFI zone.
In addition, all data in channels around 1280, 1300, 1312, 1316, and
1350\,MHz were masked as they were also completely contaminated by
satellite RFI. The resulting fraction of flagged data per channel
is shown in Fig. \ref{fig:rfi-occupancy}. 

\subsubsection{Gridding\label{sub:Gridding} }

The reduced Parkes data were gridded using \textsc{gridzilla}%
\footnote{\textsc{gridzilla} is supported by the Australia Telescope National
Facility and is available at http://www.atnf.csiro.au/computing/software/livedata/%
}. Robust median averaging was used to minimize the impact of RFI on
the final cubes and the two polarizations were added. Pixel sizes
in the gridded data are each $4\times4$\,arcmin$^{2}$. For more
details, see \citet{barnes01}. 

The SGP field is sufficiently small that it is reasonable to produce
a single cube in each frequency band. To account for the band overlap,
each cube was cut at 1310\,MHz and the two were then concatenated
together such that the frequency coverage was continuous. The rms
noise per channel in the RFI-free part of the final data cube is 8.5\,mJy\,beam$^{-1}$.
The flux density scale is calibrated against 1934-638 and Hydra A.
Calibrations are consistent to within $\sim2$ per cent. 

Since the sky coverage of the HIPASS data is so large, the southern
region is split into 388 separate sky fields, each measuring $8\times8$\,deg$ $$^{2}$
\citep{barnes01}. The northern HIPASS field consists of 102 $8\times8$\,deg$^{2}$
cubes and 48 $8\times7$\,deg$^{2}$ cubes \citep{wong06}. The rms
noise level of the HIPASS data is 13.3\ mJy \citep{barnes01}, which
is 56 per cent higher than the SGP data. The effective spatial resolution
of both the gridded SGP and HIPASS data sets is 15.5\,arcmin.

\section{Analysis\label{sec:Analysis}}

\subsection{SGP source selection\label{sub:Source-selection-sgp}}

There are 4,240 2dFGRS sources in the SGP field with high-quality
redshifts ($Q\geq3$) covered by the 21\,cm observations. However,
not all of these are suitable for use in this stacking analysis. All
channels within 2\,MHz of the extremes of the spectral range were
discarded due to the bandpass shape, and the corresponding sources
rejected. Therefore, we have a final redshift coverage of $0.0405<z<0.1319$
for this SGP sample. Additionally, sources were rejected from the
sample if their redshifted {\hi} signatures fell at frequencies with
100 per cent RFI occupancy in the data (see Fig. \ref{fig:rfi-occupancy}). 

Continuum sources must be avoided as they can induce standing waves
in the data \citep{barnes01} and create a discontinuity between the
two spectral bands. Using the NRAO VLA Sky Survey (NVSS) source catalogue
by \citet{condon98}, the 17 continuum sources in the SGP field with
1.4\,GHz flux densities greater than 200\,mJy were identified. The
840 2dFGRS galaxies with angular separations from these continuum
sources of less than the spatial resolution of the data were discarded
from the sample.

\begin{figure}
\includegraphics[bb=30bp 10bp 530bp 400bp,clip,scale=0.45]{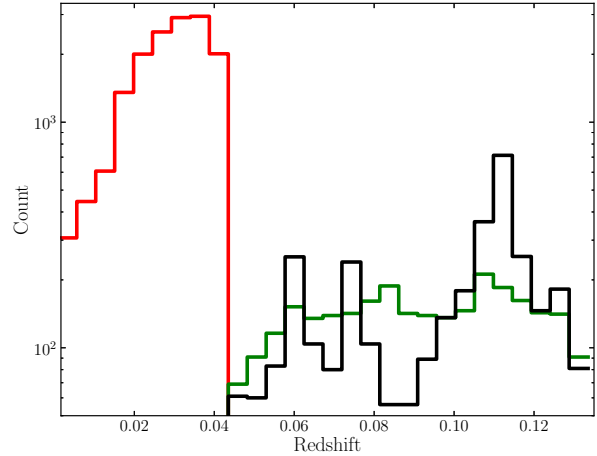}

\caption{The heliocentric redshift distributions of the optical samples considered
here. The black line shows\textbf{ }the 3,277 2dFGRS sources within
the observed SGP region and spectral range suitable for stacking.
The mean redshift is 0.096. Overlaid in green is the redshift distribution
of the full 2dFGRS catalogue across this redshift range, scaled by
the ratio of the SGP to total 2dFGRS field areas. The red line shows\textbf{
}the distribution of the 15,093 2dFGRS sources with available HIPASS
spectra. The average redshift of this sample is 0.029.}
\label{sgp_Nz}
\end{figure}

Our final 2dFGRS sample therefore contains 3,277 sources suitable
for stacking within the SGP field. The heliocentric redshift distribution
of these sources is shown in Fig. \ref{sgp_Nz}. Many galaxies are
distributed around $z=0.11$ and the average redshift of the sample
is 0.096. This `clumpiness' in redshift space is not attributable
to a single galaxy cluster, but seems to be probing genuine cosmic
structure. For example, the 2dFGRS Percolation-Inferred Galaxy Group
(2PIGG) catalogue by \citet{eke04} identifies 17 galaxy groups with
10 or more members at $z\approx0.11$ within the SGP region. 

For comparison, the redshift distribution of the full 2dFGRS catalogue
(across the relevant $z$ range) is also shown in Fig. \ref{sgp_Nz}.
It has been scaled down by the ratio of the full 2dFGRS and SGP field
areas so the shape of the two distributions can be compared. The full
2dFGRS distribution is far more uniform across redshift and the ratio
of the total source counts reveals that the SGP field contains an
$\sim30$ per cent overdensity at these redshifts. It is therefore
evident that large-scale clustering is present in the chosen SGP field,
highlighting the importance of cosmic variance considerations in such
an analysis. This is discussed in detail in Section \ref{sub:Cosmic-variance}. 

\subsection{HIPASS source selection\label{sub:Source-selection-sgp-1}}

The full HIPASS completely covers the 2dFGRS footprint, but is limited
in redshift coverage. There are 15,152 2dFGRS sources with reliable
redshifts ($Q\geq3$) within the spectral range of the HIPASS data
that are appropriate for stacking. This number excludes any sources
with $z<0.0025$ ($v<750$\ km\,s$^{-1}$) to minimize contamination
from the Galaxy and misclassified stars. We have removed the contribution
of any Galactic emission from all spectra by masking frequency channels
between 1418.4 and 1422.4\ MHz. Our final HIPASS sample consists
of 15,093 2dFGRS sources. The redshift distribution is shown in Fig.
\ref{sgp_Nz}. According to \citet{barnes01}, the contribution from
continuum sources has already been suppressed in the HIPASS data by
fitting a weighted average of the spectral shape near these sources
and subtracting from all subsequent spectra.

\subsection{Stacking process\label{sub:Stacking-process}}

\subsubsection{Extracting {\hi} spectra\label{sub:Extracting-HI-spectra}}

For each 2dFGRS source in our final sample, we extracted the full
21\,cm spectrum from the data cube using an optimal weighting according
to the beam shape, centred on the pixel corresponding to the RA and
Dec.\ defined in the optical catalogue. This extraction strategy
optimizes the S/N of the stacked spectrum but results in an expanded
effective beam width of 21.2\,arcmin ($\sim$1.6\,$h^{-1}$\,Mpc
at $z=0.1$) for the SGP data and 21.9\,arcmin ($\sim$0.6\,$h^{-1}$\,Mpc
at $z=0.03$) for the HIPASS data.

\begin{figure}
\includegraphics[bb=20bp 10bp 540bp 400bp,clip,scale=0.45]{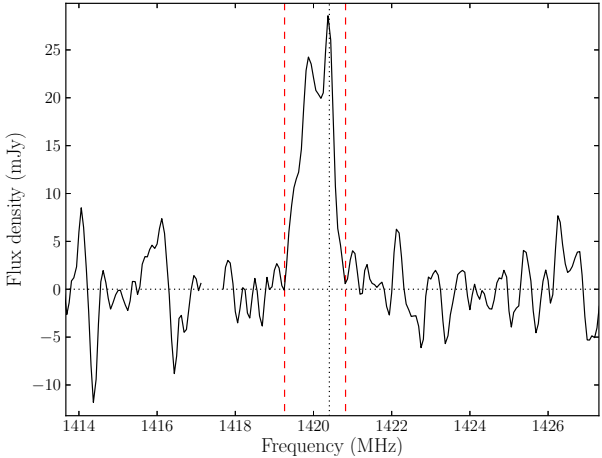}

\caption{The 21\,cm flux density spectrum of the only directly detected 2dFGRS
source in the SGP field. The gap in the spectrum is due to RFI flagging.
The dashed red lines show the nominal width of the profile and the
vertical dotted line indicates rest frequency.\label{fig:detection-spectrum}}
\end{figure}

We find one direct {\hi} detection of a 2dFGRS source in the SGP
sample with an S/N of 7.2. The {\hi} spectrum of this source is shown
in Fig. \ref{fig:detection-spectrum} and the measured parameters
are reported in Table \ref{tab:detection-parameters}. There is no
firm evidence of {\hi} detection in the rest of the sample. 

\begin{table}
\begin{centering}
\begin{tabular}{|c|c|}
\hline 
Parameter & Value\tabularnewline
\hline 
\hline 
RA (2dFGRS)  & 00:40:04.28\tabularnewline
Dec. (2dFGRS)  & -30:35:45.60\tabularnewline
$z$ (2dFGRS) & 0.0500\tabularnewline
$\nu_{\textrm{{centre}}}$ (MHz) & 1352.8\tabularnewline
$\Delta v$ (km\,s$^{-1}$) & 329\tabularnewline
S/N & 7.2\tabularnewline
$S_{{\rm int}}$ (mJy\,MHz) & 24\tabularnewline
$M$$_{\hi}$ ($\times10^{9}\, h^{-2}$\,$M$$_{\odot}$) & 29\tabularnewline
\hline 
\end{tabular}
\par\end{centering}

\caption{Measured parameters of the SGP detection in Fig. \ref{fig:detection-spectrum}.
\label{tab:detection-parameters}}
\end{table}

In contrast to this, there are numerous, significant individual detections
in the HIPASS spectra owing to the fact that HIPASS surveys the more
local Universe. The HIPASS source catalogue (HICAT) created by \citet{meyer04}
(southern field) and \citet{wong06} (northern field) identifies 5,317
{\hi} sources in total. 226 of these coincide spatially and spectrally
with 2dFGRS galaxies in our sample such that they will contribute
to the extracted {\hi} spectrum.

The {\hi} mass per unit frequency in the observed frame (hereafter
referred to as the mass spectrum $M{}_{{\rm \hi},{\color{red}{\normalcolor \nu}_{{\normalcolor {\rm obs}}}}}$)
can be calculated from: 

\noindent \begin{flushleft}
\begin{equation}
\left(\frac{{\color{red}{\normalcolor M}{}_{{\normalcolor \hi,{\color{red}{\normalcolor \nu}_{{\normalcolor {\rm obs}}}}}}}}{M_{\odot}\,{\rm {MHz}^{-1}}}\right)=4.98\times10^{7}\,\left(\frac{S_{{\color{red}{\normalcolor \nu}_{{\normalcolor {\rm obs}}}}}}{{\rm {Jy}}}\right)\,\left(\frac{D_{L}}{\textrm{Mpc}}\right)^{2}\,,\label{eq:MHI-2}
\end{equation}

\par\end{flushleft}

\noindent where \textcolor{red}{${\normalcolor S}$${\normalcolor }_{{\normalcolor \nu_{{\rm obs}}}}$
}is the observed-frame {\hi} flux density and $D$$_{L}$ is the
luminosity distance.

\subsubsection{Stacking spectra\label{subsub:Stacking-spectra}}

The first step in the stacking process is to align all extracted spectra
at rest frequency (1420.406\,MHz). The spectral axis is converted
from the observed to the emitted frame via $\nu_{{\rm em}}=\nu_{{\rm obs}}(1+z)$.
To conserve the total mass (i.e. $\int M{}_{\hi,\nu_{{\rm em}}}d\nu_{{\rm em}}=\int M{}_{\hi,\nu_{{\rm obs}}}d\nu_{{\rm obs}}$),
the {\hi} mass per unit frequency in the galaxy rest frame is given
by

\noindent \begin{flushleft}
\begin{equation}
M{}_{\hi,\nu_{{\rm em}}}=\frac{M{}_{\hi,\nu_{{\rm obs}}(1+z)}}{(1+z)}\,.\label{eq:MHI_rest_frame}
\end{equation}

\par\end{flushleft}

\noindent The data around rest frequency now contain any {\hi} emission
associated with the galaxies. These $n$ aligned spectra are then
`stacked' by finding the weighted mean in each channel:

\noindent \begin{flushleft}
\begin{equation}
\langle M_{\hi}\rangle_{\nu_{{\rm em}}}=\frac{\overset{n}{\underset{i=1}{\sum}}(w_{i}{\color{red}{\normalcolor \textrm{\ensuremath{M}\ensuremath{{}_{\hi,{\color{red}{\normalcolor \nu_{{\rm em}}}},i}}}}})}{\overset{n}{\underset{i=1}{\sum}}w_{i}}\,,\label{eq:MHI-stack}
\end{equation}

\par\end{flushleft}

\noindent where\textcolor{red}{{} ${\color{red}{\normalcolor M}}{\normalcolor }_{{\normalcolor \hi,\nu_{{\rm em}},i}}$}
is the value of the $i$th mass spectrum at emitted frequency \textcolor{red}{${\normalcolor \nu}_{{\normalcolor {\rm em}}}$}.
We choose the weighting factor $w_{i}=\sigma{}_{i}^{-2}$ (where $\sigma$
is the rms noise level of each observed-frame flux density spectrum),
which improves the S/N of the final co-added spectrum. Using $w_{i}=(\sigma_{i}D_{i}^{2})^{-2}$
would further optimize the S/N but would reduce the effective volume
and increase cosmic variance in our results.

\begin{figure*}
\includegraphics[bb=20bp 5bp 530bp 410bp,clip,scale=0.45]{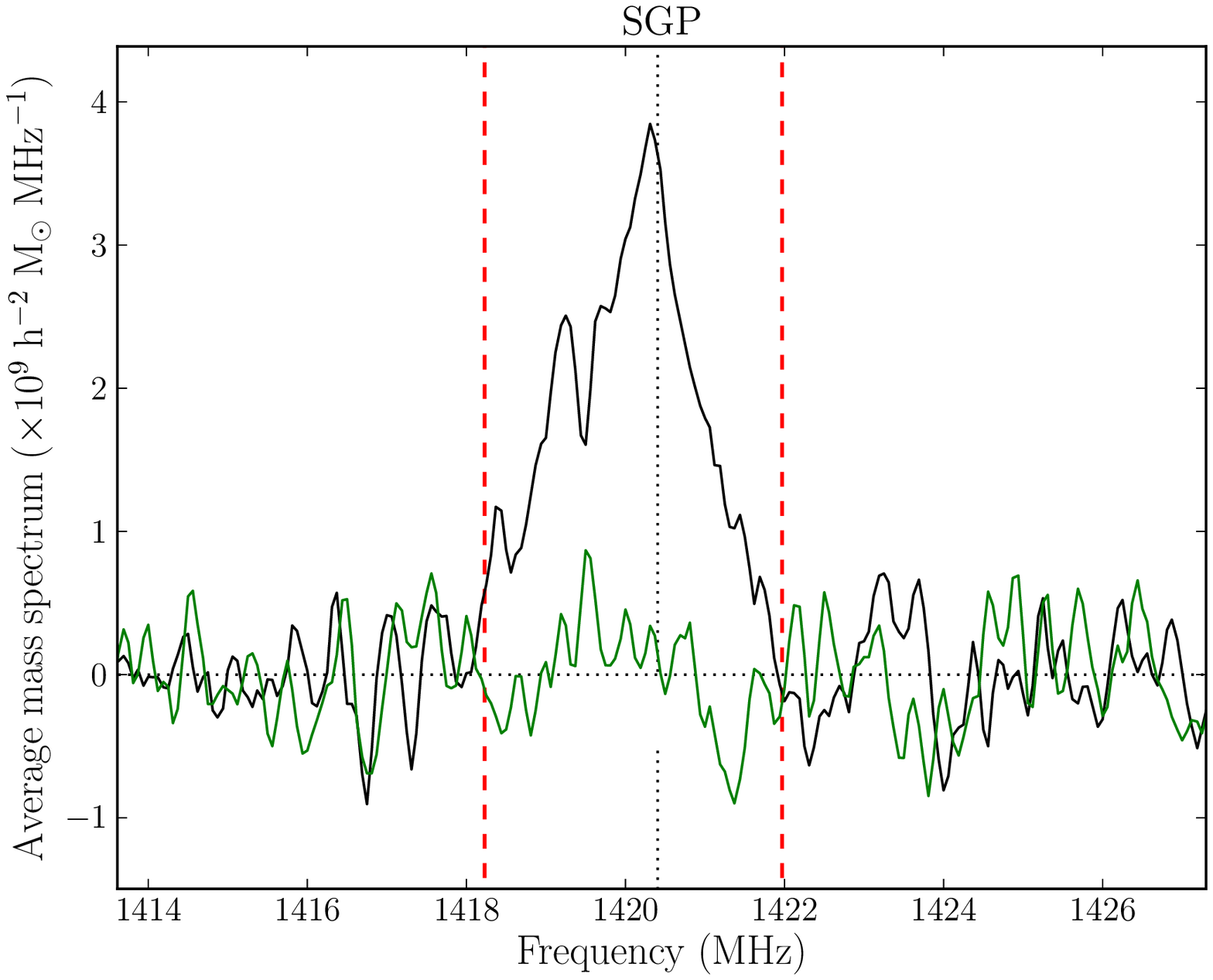}\includegraphics[bb=15bp 5bp 530bp 410bp,clip,scale=0.45]{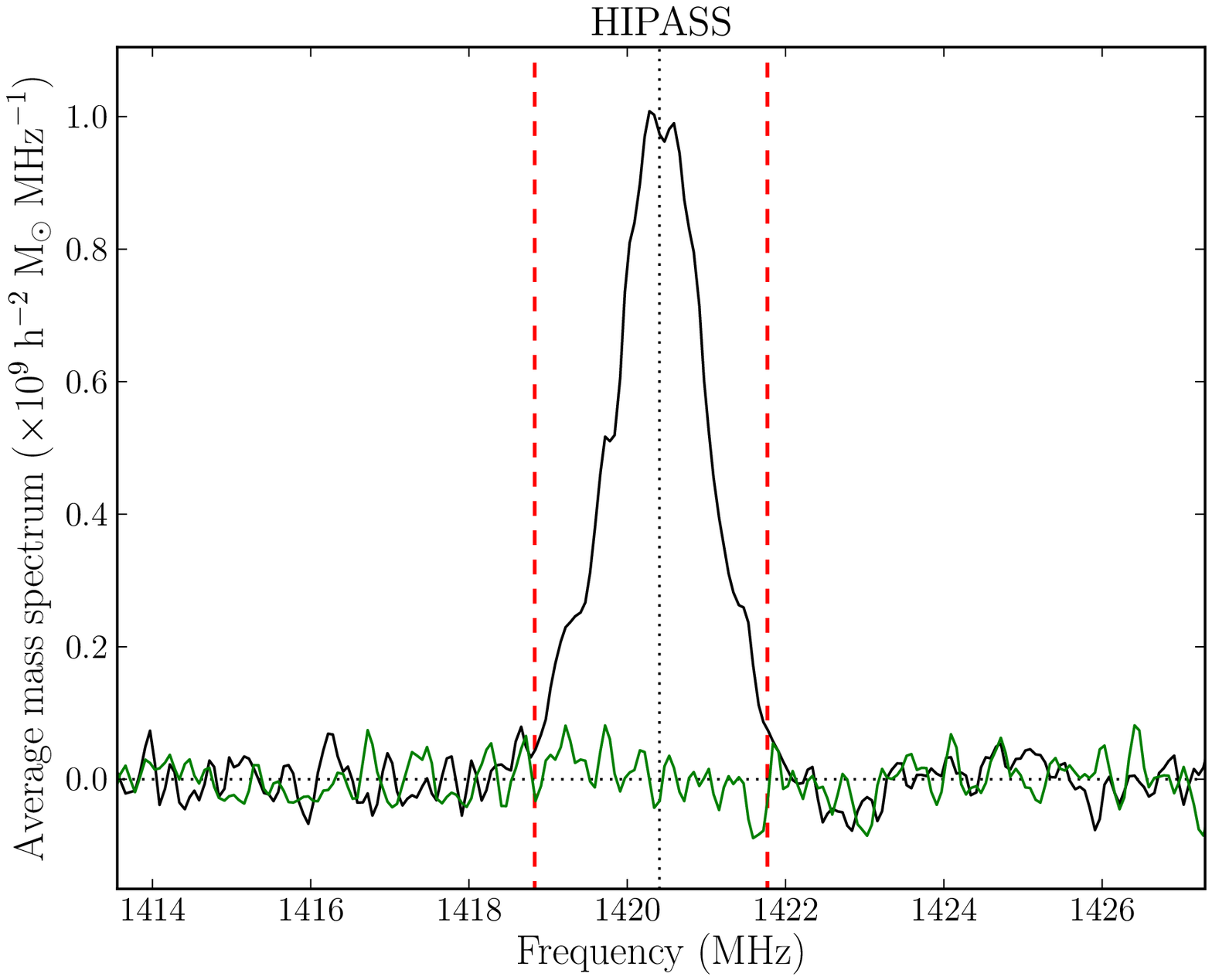}

\caption{The co-added {\hi} mass spectrum of all 3,277 SGP sources (left)
and 15,093 2dFGRS sources covered by HIPASS (right). The vertical
dotted line indicates the rest frequency of the {\hi} line. The dashed
red lines represent the nominal width of the stacked signal. The stacked
spectrum generated using the control catalogue is shown in green to
indicate the noise level in each case. A fourth-order polynomial baseline
has been subtracted from all co-added spectra. Data between the dashed
red lines were excluded when fitting baselines to the black spectra.}
\label{Flo:sgp_stacks}
\end{figure*}

The stacked spectra produced by co-adding all 3,277 SGP sources and
15,093 HIPASS sources are shown in Fig. \ref{Flo:sgp_stacks}. Although
only one galaxy is individually detected in the SGP data, a strong
12\,$\sigma$ averaged detection is achieved through stacking. As
expected, a more significant 31\,$\sigma$ stacked detection is achieved
for the HIPASS sample due to the presence of many strong individual
detections and a larger overall sample size. 

The stacked SGP profile nominally spans the spectral range indicated
by the dashed lines in Fig. \ref{Flo:sgp_stacks} and is $1418.23-1421.97$\,MHz
($\Delta v=790$\,km\,s$^{-1}$). The stacked HIPASS profile spans
$1418.83-1421.77$\,MHz ($\Delta v=621$\,km\,s$^{-1}$). These
are significantly wider than the expected {\hi} line widths of the
individual galaxies contributing to the stack (see Section \ref{sub:Line-width-estimation}).

Possible explanations for the broadened stacked profiles include:
(i) systematic differences between optical and radio redshifts, (ii)
errors in the optical redshifts (85\,km\,s$^{-1}$) and (iii) source
confusion. The principle cause is likely confusion, particularly in
the case of the higher redshift SGP data where a larger volume is
probed per beam. This will be explored in detail in Section \ref{sub:Confusion}.

Note that the HIPASS spectrum is symmetric as expected, yet the SGP
spectrum is $\sim$0.62\,MHz ($130$\,km\,s$^{-1}$) wider at frequencies
below the rest frame. This does not appear to be related to optical
redshift accuracy.

The green line in Fig. \ref{Flo:sgp_stacks} indicates the noise level
from averaged mock spectra created by stacking sources from a control
catalogue. This catalogue was generated by randomly mismatching 2dFGRS
redshifts and the source positions. By using the false redshifts in
the extraction process, the channels shifted to 1420.4\,MHz should
not contain any {\hi} emission. Thus, co-adding these mock spectra
will not produce any stacked detection but will closely approximate
the noise behaviour. A baseline was removed from all stacked spectra
via fourth-order least-squares polynomial fitting over a 14\,MHz
spectral range, excluding the signal region.\textbf{ }

\subsubsection{Average {\hi} mass\label{sub:Average-HI-mass}}

Integrating over the channels containing the stacked detection (${\color{red}{\normalcolor \nu}_{{\normalcolor {\rm em},1}}}$
to ${\color{red}{\normalcolor \nu}_{{\normalcolor {\rm em},2}}}$)
gives the average {\hi} mass of the sample:

\begin{equation}
{\color{red}{\normalcolor \left\langle M_{{\rm \hi}}\right\rangle =\int_{{\color{red}{\normalcolor \nu}_{{\normalcolor {\rm em},1}}}}^{{\color{red}{\normalcolor \nu}_{{\normalcolor {\rm em,}2}}}}\left\langle {\color{red}{\normalcolor M}{}_{{\rm {\normalcolor \hi}}}}\right\rangle _{{\color{red}{\normalcolor \nu}_{{\normalcolor {\rm em}}}}}d{\color{red}{\normalcolor \nu}_{{\normalcolor {\rm em}}}}\,.}}\label{eq:mean MHI}
\end{equation}

We find average {\hi} masses of $\langle M_{\hi}\rangle=(6.93\pm0.17)\times10^{9}$\,$h^{-2}$\,${\rm M}{}_{\odot}$
for the SGP sample and $\left\langle M_{{\rm \hi}}\right\rangle =(1.48\pm0.03)\times10^{9}\, h^{-2}$\,${\rm M}_{\odot}$
for the HIPASS sample. 

The error on the average mass is estimated by producing 10 control
catalogues as in Section \ref{subsub:Stacking-spectra}, repeating
the stacking process for each, and then calculating the rms deviation
in the integrated mass over the signal region.

In both cases, we appear to be sampling galaxies with average masses
of less than $M_{\hi}^{*}$=$8.4\times10^{9}\, h^{-2}\,{\rm M}_{\odot}$
\citep{zwaan05}. The 5\,$\sigma$ detection limit at the mean redshift
of the SGP sample corresponds to an {\hi} mass of $\sim1\times10^{11}$\,$h^{-2}\,{\rm M}_{\odot}$.

Note, however, that the average masses calculated here will be larger
than the true value for each sample. This is due to significant source
confusion in the data and will be discussed in Section \ref{sub:Confusion}.
The significantly smaller mass obtained for the HIPASS sample is due
to the 2dFGRS magnitude limit which allows more nearby, low-luminosity
galaxies to appear, and due to less confusion with other galaxies
in the telescope beam.

\section{Results}

\subsection{Noise behaviour\label{sub:Noise-Behaviour}}

The best possible S/N achievable with a stacking analysis will be
obtained if the noise behaves in a purely Gaussian manner and decreases
with the square root of the number of co-added spectra. To demonstrate
the noise behaviour of the Parkes data, we stack $N$ randomly selected
spectra taken from the control catalogues described in Sections \ref{subsub:Stacking-spectra}
and \ref{sub:Average-HI-mass}. Note that in this case we are stacking
the flux density spectra, as opposed to the mass spectra. We then
find the rms noise per channel calculated over the spectral range
indicated by the dashed lines in Fig. \ref{Flo:sgp_stacks}.

\begin{figure}
\noindent \begin{raggedright}
\includegraphics[bb=10bp 0bp 530bp 400bp,clip,scale=0.45]{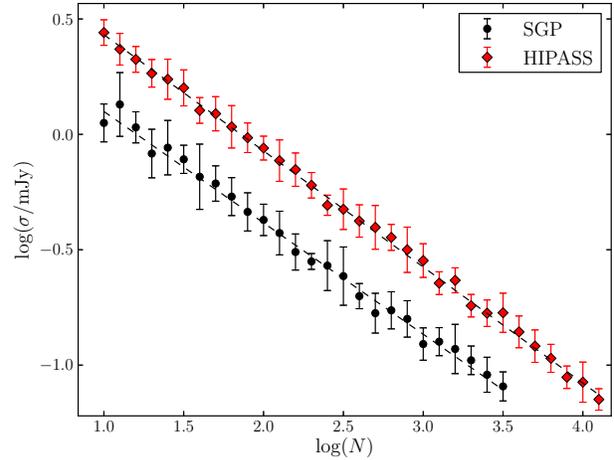}
\par\end{raggedright}

\caption{The rms noise level of the co-added spectrum versus the number of
individual spectra contributing to the stack. The round black points
are generated using SGP data and the red diamonds with HIPASS. The
dashed lines show the best fit to the data. Both have gradients close
to -0.5. }
\label{Flo:NvN}

\end{figure}

Fig. \ref{Flo:NvN} shows the noise level of the stacked spectrum
as a function of the number of co-added spectra. It is evident that
the noise within the signal region of the stacked 21\,cm spectra
in both data sets displays Gaussian behaviour, with gradients $-0.48\pm0.05$
(SGP sample) and $-0.50\pm0.03$ (HIPASS sample). This demonstrates
that our data reduction has been successful in mitigating against
non-Gaussian noise contributions from eg. RFI, residual continuum
emission and insufficient bandpass calibration. If we do not apply
these careful corrections to the data, we see evidence of a noise
`floor,' i.e. the sensitivity does not continue to decrease with the
expected gradient, but flattens out once the non-Gaussian components
begin to dominate. 

We reach a final sensitivity of 68\,$\mu$Jy by stacking all 3,277
SGP spectra and 68\,$\mu$Jy by stacking all 15,093 HIPASS spectra.
The integration time is 975\,s/pixel in the SGP data and 220\,s\,pixel$^{-1}$
in HIPASS \citep{barnes01}. This difference explains the vertical
displacement of the SGP and HIPASS noise trends in Fig. \ref{Flo:NvN}.
Despite the greater number of sources contributing to the final HIPASS
stack, the longer integration time per pixel for the SGP data results
in a similar final sensitivity. Observation of a single source for
\textbf{$\sim37$} d would be required to attain the same sensitivity
as these final stacked spectra.

Note, however, that the mass spectra in Fig. \ref{Flo:sgp_stacks}
incorporate a $D_{L}$$^{2}$ factor (see equation \ref{eq:MHI-2}),
which is larger for the higher redshift SGP sample. The rms noise
level in the stacked SGP mass spectrum is therefore six times larger
than for HIPASS.

\subsection{Confusion\label{sub:Confusion}}

The accuracy of the average quantities\textbf{ }we measure via this
stacking method will ultimately be limited by confusion in the Parkes
data. We are unable to distinguish between multiple objects within
the Parkes beam and therefore additional {\hi} flux from neighbouring
galaxies can contaminate our measurements.

\noindent Based on the simulated results of \citet{obreschkow13}
and the observed {\hi} line widths of HICAT galaxies, it is reasonable
to assume a maximum line width of 600\,km\,s$^{-1}$ for any individual
galaxy contributing to our analysis.

As discussed in Section \ref{subsub:Stacking-spectra} above, the
stacked {\hi} signals we achieve are broader than this, particularly
for the SGP sample. Therefore, we consider all the flux in the wings
of the stacked {\hi} profile, beyond $\pm$300\,km\,s$^{-1}$ ($\pm$1.42
MHz) from the rest frame, to exist solely due to confusion with other
galaxies in the 2dFGRS catalogue, as well as fainter sources below
its magnitude limit. 

Restricting the integral in equation \prettyref{eq:mean MHI} to this
spectral range would partially correct for confusion in the measured
$\langle M_{{\rm \hi}}\rangle$. However, any flux within $\pm300$\,km\,s$^{-1}$
will still include contributions from confused sources, as well as
from the target galaxies themselves. The challenge is therefore to
identify which sources are confused over this spectral range, and
then to correct for this.

\subsubsection{Line width estimation with the Tully-Fisher relation\label{sub:Line-width-estimation}}

We do not directly detect the {\hi} signature of many galaxies in
our sample and so cannot determine their {\hi} profile widths. Therefore,
we cannot ascertain in advance whether we are integrating over single
or multiple {\hi} profiles and hence we cannot easily discern the
extent of spectral confusion.

One way to overcome this is to estimate the {\hi} line widths of
each galaxy in our sample using the Tully-Fisher relation. We choose
the $B$-band relation of \citet{meyer08} which was derived using
optical counterparts to HIPASS galaxies:

\begin{equation}
\log(v_{{\rm {rot}}})=-(M_{B}+5\log h+1.4)/8.6\,,\label{eq:TF}
\end{equation}

\noindent where $v_{{\rm {rot}}}$ is the rotational velocity of the
galaxy (in km\,s$^{-1}$) and $M_{B}$ is the $B$-band absolute
magnitude. $B$ apparent magnitudes are estimated from the SuperCOSMOS
$b_{J}$ and $r_{F}$ apparent magnitudes available in the 2dFGRS
catalogue as follows%
\footnote{Derived from magnitude conversions available at www2.aao.gov.au/2dfgr/%
}:

\begin{equation}
B=b_{J}-0.0047+0.236(b_{J}-r_{F})\,.\label{eq:bj-to-B}
\end{equation}

\begin{figure}
\includegraphics[bb=15bp 5bp 535bp 400bp,clip,scale=0.45]{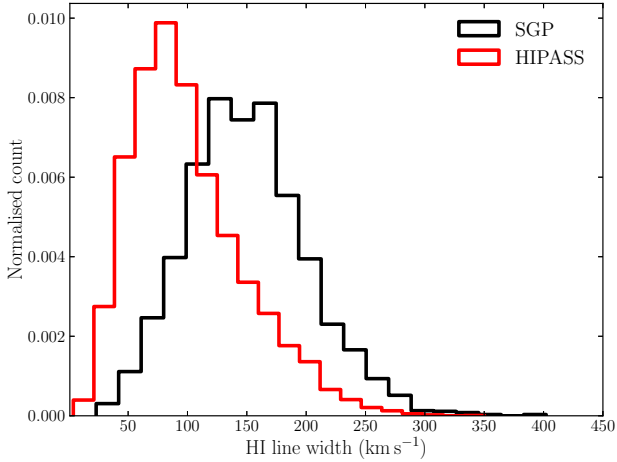}

\caption{{\hi} line widths estimated using the Tully-Fisher relation for the
sample of SGP (black) and HIPASS (red) galaxies. Quantities have been
normalized to show the relative spread in estimated line widths.\label{fig:LW}}

\end{figure}

\noindent SuperCOSMOS magnitudes used in equation \prettyref{eq:bj-to-B}
have not been corrected for extinction by the Galaxy. However, we
only require indicative values for $B$ and $v_{{\rm rot}}$. The
{\hi} line width $W_{{\rm \hi}}$ (defined at 50 per cent peak flux)
can then be estimated using:

\begin{equation}
W_{{\rm \hi}}=2v_{{\rm {rot}}}\sin(i)\,,\label{eq:LW}
\end{equation}

\noindent where $i$ is the inclination of the galaxy, calculated
from the observed axial ratio as in equation (24) in \citet{meyer08}.
The distribution of the resulting line width estimates is shown in
Fig. \ref{fig:LW}. The average width is 150\,km\,s$^{-1}$ for
SGP and 100\,km\,s$^{-1}$ for HIPASS. This seems reasonable in
comparison to the {\hi} line widths of HICAT galaxies, which have
an average of 178\,km\,s$^{-1}$.

\subsubsection{Identification of confused sources}

Using the Tully-Fisher predicted line widths, we can now identify
which 2dFGRS galaxies are likely to give rise to confusion over the
spectral range of interest. We consider a source confused with a particular
target galaxy if: (i) it is within one beam width of the target galaxy
($\pm$21.2\,arcmin for the SGP, $\pm21.9$\,arcmin for HIPASS)
and (ii) if any part of its predicted {\hi} profile is within $\pm$300\,km\,s$^{-1}$
of the target galaxy's recessional velocity. 

As expected, we find significantly more confusion in the higher redshift
SGP field. On average, any one SGP galaxy is confused with seven others.
In the lower redshift HIPASS data, this decreases to three. 

\subsubsection{Luminosity correction\label{sub:Luminosity-correction}}

The goal of this paper is to derive an estimate of {\hi} density,
based on the mass-to-light ratio of the galaxies. While we cannot
fully correct $\langle M_{{\rm \hi}}\rangle$ for confusion, we can
account for this in the density estimate by artificially `confusing'
the optical luminosities. We do this by defining

\begin{equation}
{\textstyle L'=\overset{n}{f\underset{i=1}{\sum}}L_{i}b_{i}}+L_{0}\,,\label{eq:lum_corr}
\end{equation}

\noindent where $L{}_{0}$ is the luminosity of the central galaxy,
$L_{i}$ are the luminosities of the $n$ confused galaxies, $b_{i}$
is the beam weighting factor defined using a Gaussian with full width
at half maximum (FWHM)$=$21.2\,arcmin (SGP) and FWHM$=$21.9\,arcmin
(HIPASS), and $f$ is an extrapolation factor to account for luminosity
from galaxies fainter than the 2dFGRS survey limit. The latter assumes
the luminosity density function of \citet{norberg02}, accounting
for evolution and $k$-corrections.

\begin{figure*}
\includegraphics[bb=30bp 0bp 540bp 415bp,clip,scale=0.45]{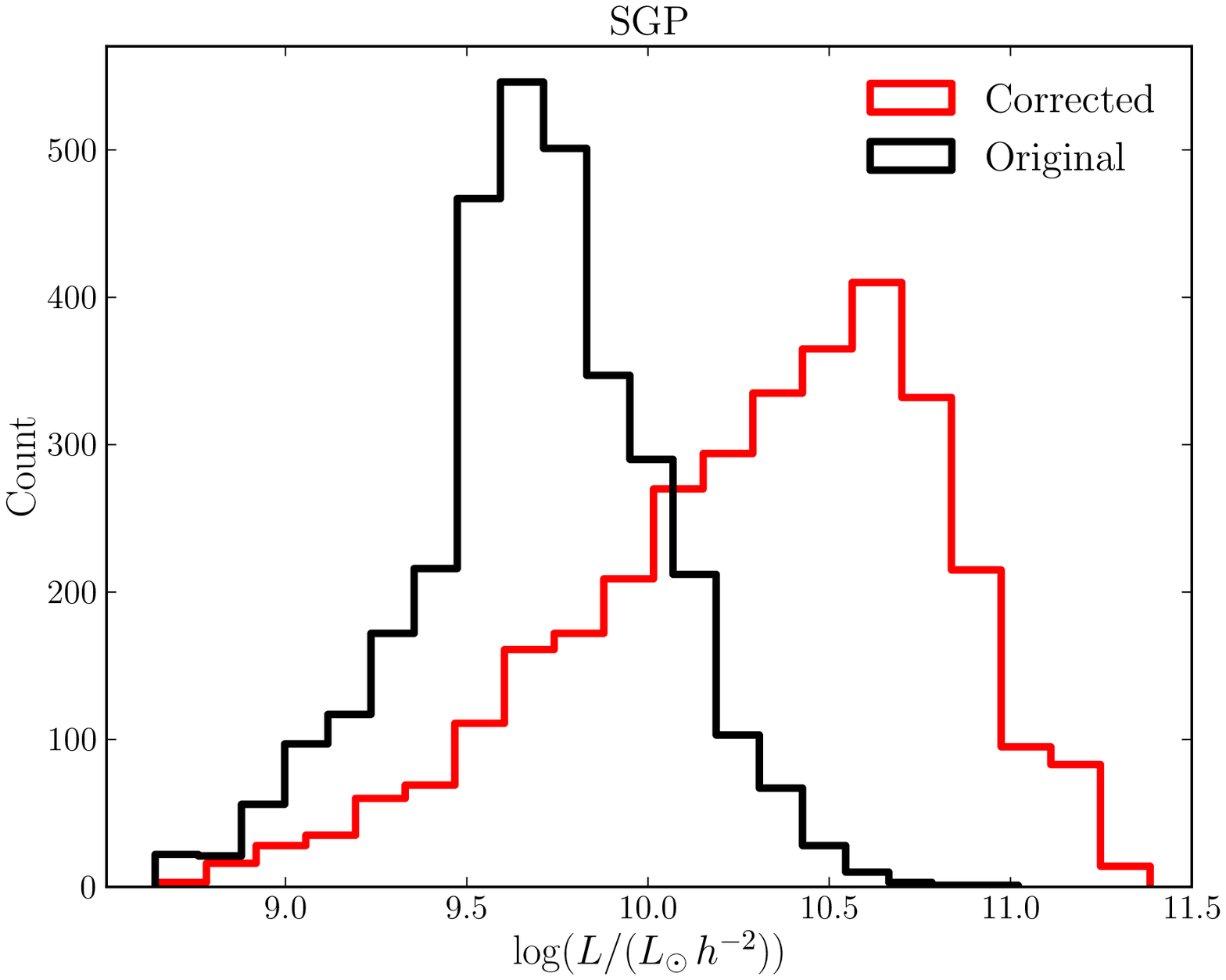}\includegraphics[bb=20bp 0bp 540bp 415bp,clip,scale=0.45]{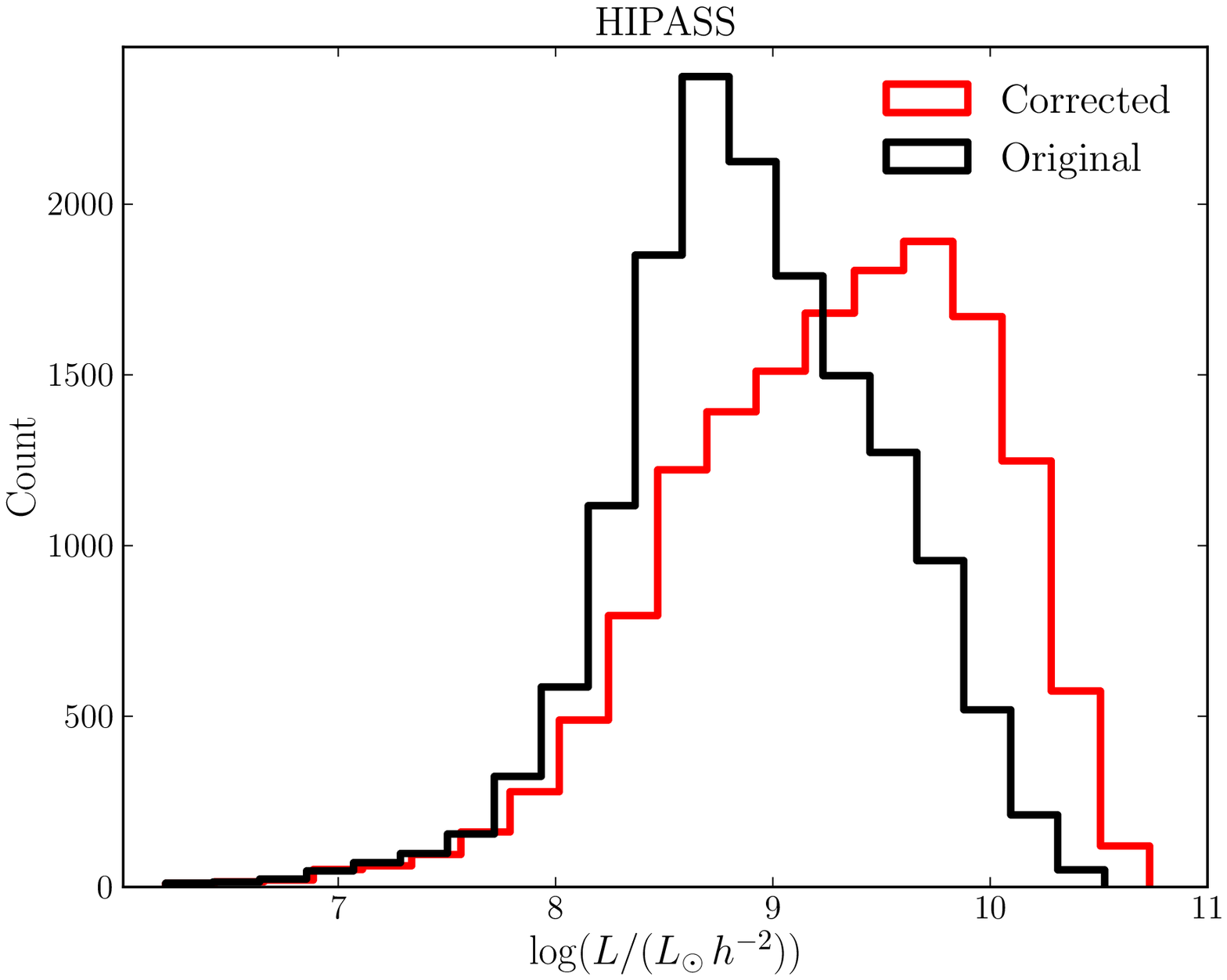}

\caption{The luminosity distribution of all stacked optical sources in the
SGP field (left) and HIPASS field (right). The black line represents
the original b$_{J}$ luminosities of the sample, as defined in the
2dFGRS source catalogue. The red line shows the distribution once
luminosities have been adjusted to include a beam-weighted sum of
the luminosities of all confused galaxies.}
\label{Flo:Ldistbn}
\end{figure*}

Fig. \ref{Flo:Ldistbn} shows the original $b$$_{J}$ luminosity
distribution of each sample, compared to the confusion-adjusted luminosity
distribution ($L'$). The average (catalogued) luminosity for the
SGP sample is $6.62\times10^{9}\, h^{-2}\,{\rm L}_{\odot}$. Once
adjustment for confusion has been carried out, this average increases
by a factor of 5 to $3.36\times10^{10}\, h^{-2}\,{\rm L}_{\odot}$.
The average original luminosity of the HIPASS sample is $1.93\times10^{9}\, h^{-2}\,{\rm L}_{\odot}$,
which increases by a factor of 2.5 to $4.87\times10^{9}\, h^{-2}\,{\rm L}_{\odot}$
when adjusted. As expected, the correction to the HIPASS luminosities
is less significant to that of the SGP, due to less source confusion
and a fainter absolute luminosity limit.

\subsection{{\hi} mass density\label{sub:Cosmic-HI-mass}}

The {\hi} density, $\rho_{{\rm \hi}}$, can be calculated from the
luminosity density and the mass-to-light ratio using

\noindent \begin{center}
\begin{equation}
\rho_{\hi}=\left\langle \frac{M_{\hi}}{L}\right\rangle _{\rho_{L}}\times\rho_{L}\,,\label{eq:rhohi}
\end{equation}

\par\end{center}

\noindent as in eg. \citet{fall93}. We use the $b_{J}$ luminosity
density $\rho_{L}=(1.82\pm0.17)\times10^{8}\, h$\,$L$$_{\odot}$\,Mpc$^{-3}$
derived by \citet{norberg02} using the full 2dFGRS catalogue, with
appropriate evolution and $k$-corrections applied (up to 17 per cent
at the redshifts of the SGP galaxies). 

\noindent We calculate $\langle M_{\hi}/L\rangle$ by stacking individual
$M_{\hi}/L$ `spectra'. We do so via equations \prettyref{eq:MHI-stack}
and \prettyref{eq:mean MHI}, replacing $\ensuremath{M}{}_{\hi,{\color{red}{\normalcolor \nu}_{{\normalcolor {\rm em}}}},i}$
with ${\normalcolor {\color{red}{\normalcolor M}{\normalcolor _{\hi,{\color{red}{\normalcolor \nu}_{{\normalcolor {\rm em}}}},i}}}}/L'_{i}$.
As discussed in Section \ref{sub:Confusion}, it is appropriate to
restrict the spectral integration range to $\pm300$\,km\,s$^{-1}$
from the rest frame since any contribution outside of this is due
only to confused sources. We find $\langle M_{\hi}/L\rangle=0.43\pm0.03$\,${\rm M}_{\odot}/{\rm L}_{\odot}$
for the SGP sample, and $0.72\pm0.03$\,${\rm M}_{\odot}/{\rm L}_{\odot}$
for the HIPASS sample, where the errors are measured as in Section
\ref{sub:Average-HI-mass}. 

We require the mass-to-light ratio to be weighted by the luminosity
density, rather than the luminosity selection function of the input
catalogue (hence the subscript $\rho_{L}$ in equation \ref{eq:rhohi}).
We therefore apply a weighting correction to the mean mass-to-light
ratio as follows:

\noindent \begin{center}
\begin{equation}
\left\langle \frac{M_{\hi}}{L'}\right\rangle _{\rho_{L}}=\left\langle \frac{M_{\hi}}{L'}\right\rangle _{{\rm s}}\times\frac{\int N(L)dL}{\int L^{\alpha}N(L)dL}\times\frac{\int L^{\alpha}\rho_{L}(L)dL}{\int\rho_{L}(L)dL}\,,\label{eq:ML_weight}
\end{equation}

\par\end{center}

\noindent where $\rho_{L}(L)$ is the luminosity density function
and $N(L)$ is the original luminosity distribution of the input redshift
sample. The parameter $\alpha$ is the power-law dependence of the
average {\hi} mass-to-light ratio of galaxies on luminosity in solar
units:

\begin{equation}
\frac{M_{{\rm \hi}}}{L}=\beta L^{\alpha}\,.\label{eq:ML_trend}
\end{equation}

\noindent We adopt the $z=0$ values $\alpha=-0.38$ and $\log\beta=(3.34-2\alpha\log h$)
found by \citet{karachentsev08} using a subset of HIPASS galaxies
with optical counterparts. For the HIPASS sample, the weighting factor
applied to the mean mass-to-light ratio is 0.62. For the SGP sample,
where galaxies are more luminous and have lower {\hi} mass-to-light
ratios, it is 1.35. That is, the mean {\hi} mass-to-light ratio is
predicted to be 2.2 times greater for galaxies in the HIPASS sample
compared with the SGP sample on the basis of our previous knowledge
of the luminosity dependence of $M_{\hi}/L$.

We convert $\rho_{\hi}$ into a fraction of the critical density of
the Universe per unit comoving volume, $\Omega_{\hi}$, using 

\noindent \begin{center}
\begin{equation}
\Omega_{{\rm \hi}}=\frac{8\pi G\rho_{{\rm \hi}}}{3H_{0}^{2}}\,,\label{eq:ohi}
\end{equation}

\par\end{center}

\noindent where $G$ is the gravitational constant and $H_{0}$ is
the Hubble constant. We find $\Omega_{\hi}=(3.19_{-0.59}^{+0.43})\times10^{-4}\, h^{-1}$
for the $0.04<z<0.13$ SGP data and $(2.82{}_{-0.59}^{+0.30})\times10^{-4}\, h^{-1}$
for the $z<0.04$ HIPASS data. This implies that there has been no
(12$\pm$23 per cent) evolution in the cosmic {\hi} mass density
over the past 1.2\,$h^{-1}$\,Gyr.

The errors in $\Omega_{\textrm{\hi}}$ are a combination of: (i) the
\textbf{$\sim$}6 per cent error in $\langle M_{\hi}/L\rangle$ quoted
previously, (ii) a 9 per cent error in $\rho_{L}$ and (iii) the 1$-$11
per cent estimated uncertainty in the confusion correction. The latter
was estimated from: (a) variance in $\langle M_{\hi}/L\rangle$ in
different windows from $\pm200$ to $\pm400$\,km\,s$^{-1}$ over
the stacked spectrum and (b) variance in $\langle M_{\hi}/L\rangle$
calculated using different angular resolutions (and corresponding
confusion corrections) in the range $15.5-21.9$ arcmin.

\begin{figure}
\includegraphics[bb=15bp 0bp 530bp 395bp,clip,scale=0.48]{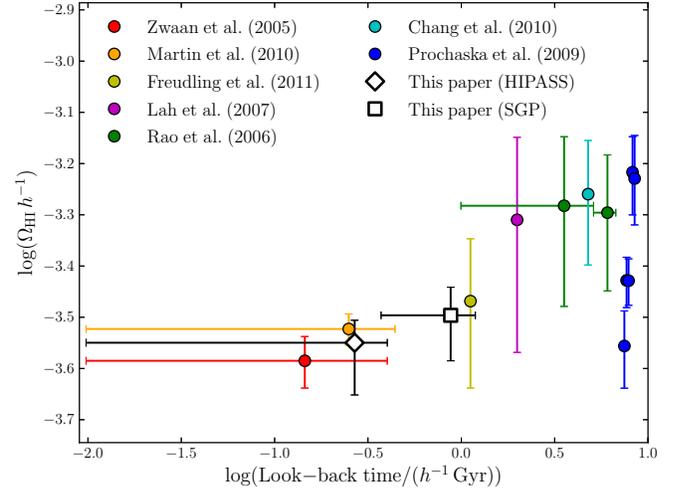}

\caption{Observational constraints on the cosmic {\hi} mass density $\Omega_{\textrm{\hi}}$
as a function of look-back time. The horizontal error bars indicate
the full redshift range over which the measurement applies. The $\Omega_{\hi}$
estimates using spectral stacking of HIPASS and SGP data are shown
as the black diamond and square points, respectively.}
\label{Flo:omegahi}
\end{figure}

Table \ref{tab:OHI values} and Fig. \ref{Flo:omegahi} compare the
values calculated here to other observational constraints on $\Omega_{\textrm{\hi}}$
from the literature. Values at look-back times of less than 4\,$h^{-1}$\,Gyr
have been calculated from either direct or stacked 21\,cm detections.
The \citet{chang10} point at 4.8\,$h^{-1}$\,Gyr was derived using
21\,cm intensity mapping. All other points estimate the {\hi} column
density of galaxies from the absorption spectra of background QSOs.
The evolutionary trends in Fig. \ref{Flo:omegahi} are flatter than
other examples of this plot in the literature because we have removed
the contribution of helium from the values presented in \citet{rao06}
and \citet{lah07} and used the latest DLA measurements of \citet{prochaska09}
rather than \citet{prochaska05}.

The $\Omega_{\textrm{\hi}}$ values we derive here are consistent,
within the error margins, with previous estimates over the same redshift
range by \citet{zwaan05}, \citet{martin10} and \citet{freudling11}.
It is particularly interesting to note the close agreement between
our low-redshift result, derived by stacking HIPASS detections and
non-detections, and that of \citet{zwaan05}, derived from only direct
HIPASS detections.

It is possible that we have underestimated our uncertainties in the
confusion correction. Perhaps, due to evolution, there are more low-mass
galaxies in our beam than we have accounted for. Whether or not this
is the case will be apparent in future deep studies with radio telescopes
with smaller beam sizes. However, it could also be that the other
measurements listed have missed galaxies, perhaps because of the difficulty
in defining selection functions or extrapolating the {\hi} density
function, and are not as accurate as claimed. The use of the stacking
method means that the contribution from even low {\hi} mass galaxies
is included. Furthermore, as will be apparent in the next section,
previous surveys may in fact be completely dominated by cosmic variance. 

\begin{table*}
\begin{tabular}{cccc}
\hline 
\multicolumn{1}{c|}{Source} & \multicolumn{1}{c|}{$\langle z\rangle$} & \multicolumn{1}{c|}{Look-back time ($h^{-1}$\,Gyr)} & $\Omega_{\textrm{\hi}}$ $(\times10^{-4}\, h^{-1})$\tabularnewline
\hline 
\hline 
\citet{zwaan05} & 0.015 & 0.15 & $2.6\pm0.3$\tabularnewline
\textbf{This paper (HIPASS)} & $\mathbf{0.028}$ & \textbf{$\mathbf{0.27}$} & $\mathbf{2.82}\mathbf{}_{\mathbf{-0.59}}^{\mathbf{+0.30}}$\tabularnewline
\citet{martin10} & $0.026$ & 0.25 & $3.0\pm0.2$\tabularnewline
\textbf{This paper (SGP)} & $\mathbf{0.096}$ & $\mathbf{0.88}$ & $\mathbf{3.19{}_{-0.59}^{+0.43}}$\tabularnewline
\citet{freudling11} & 0.125 & 1.12 & $3.4\pm1.1$\tabularnewline
\citet{lah07}$^{a}$ & 0.24 & 1.99 & $4.9\pm2.2$\tabularnewline
\citet{rao06}$^{a,b}$ & 0.505 & 3.55 & $5.2\pm1.9$\tabularnewline
\citet{chang10} & $0.80$ & 4.78 & $5.5\pm1.5$\tabularnewline
\citet{rao06}$^{a,b}$ & 1.275 & 6.07 & $5.0\pm1.5$\tabularnewline
\citet{prochaska09}$^{b}$ & 2.31 & 7.47 & $2.78_{-0.48}^{+0.48}$\tabularnewline
\citet{prochaska09}$^{b}$ & 2.57 & 7.68 & $3.73_{-0.43}^{+0.41}$\tabularnewline
\citet{prochaska09}$^{b}$ & 2.86 & 7.87 & $3.73_{-0.39}^{+0.39}$\tabularnewline
\citet{prochaska09}$^{b}$ & 3.22 & 8.06 & $5.24_{-0.46}^{+0.44}$\tabularnewline
\citet{prochaska09}$^{b}$ & 3.70 & 8.26 & $6.07_{-1.06}^{+1.05}$\tabularnewline
\citet{prochaska09}$^{b}$ & 4.39 & 8.48 & $5.90_{-1.11}^{+1.26}$\tabularnewline
\hline 
\end{tabular}

\caption{Observational constraints on $\Omega_{{\rm \hi}}$ from the literature.
All values have been converted to the same cosmology. The mean redshift
and associated look-back time of each sample are quoted. Results presented
in this paper are shown in bold. Errors on our values include statistical
uncertainty in our stacking analysis, uncertainty in the 2dFGRS luminosity
density and systematic errors introduced by correction for confusion.
$^{a}$ Original values quoted were $\Omega_{{\rm gas}}$ and included
a He contribution but have been converted to $\Omega_{\hi}$ here.
(See source for the value of the assumed contribution). $^{b}$ Values
were calculated using only systems with {\hi} surface densities $>2\times10^{20}$\,cm$^{-2}$
(i.e. measures $\Omega_{\hi}^{{\rm DLA}}$). \label{tab:OHI values}}
\end{table*}

\subsection{Cosmic variance\label{sub:Cosmic-variance}}

When observations are limited to a finite volume of the Universe,
a significant contribution to the error in estimates of cosmological
values, such as $\Omega_{\hi}$, can come from cosmic variance. That
is, variation in galaxy density due to large-scale structure. By employing
the stacking technique, we are not restricted to volumes where direct
detections are possible. Therefore, the volumes we probe are $3.7\times10^{5}$$ $
and $3.4\times10^{5}\, h^{-3}$\,Mpc$^{3}$ for the SGP and HIPASS
samples, respectively. Within the median 2dFGRS redshift range ($z<0.11$),
the SGP volume reduces to $1.8\times10^{5}\, h^{-3}$\,Mpc$^{3}$.
These are significantly larger than the sampled volumes of \citet{zwaan05}
using HICAT ($8.2\times10^{4}\, h^{-3}$\,Mpc$^{3}$, for a median
redshift of 0.009), \citet{martin10} using the 40 per cent ALFALFA
source catalogue ($1.4\times10^{5}\, h^{-3}$\,Mpc$^{3}$, for a
median redshift of 0.027) and \citet{freudling11} using the AUDS
pilot survey ($7.5\times10^{2}\, h^{-3}$\,Mpc$^{3}$). Furthermore,
the luminosity density $\rho_{L}$ is calculated by \citet{norberg02}
from an even higher volume of $7.3\times10^{6}\, h^{-3}$\,Mpc$^{3}$.
Therefore, uncertainty in $\Omega_{\hi}$ due to cosmic variance is
dramatically lower in our results than in than any previous sample.

As discussed in Section \ref{sub:Source-selection-sgp}, we have chosen
a region which is not representative of the average Universe at $0.04<z<0.13$.
Rather, the SGP field contains a small overdensity. The selection
of an over-dense region gives a higher volume-integrated {\hi} mass.
However, to first order, this should not bias our measurement of the
average mass-to-light ratio of galaxies, and therefore the {\hi}
mass density we calculate will be a fair estimate of the cosmic value.
While the local environment is known to influence the {\hi} properties
of galaxies (eg. \citealt{haynes84}), little is known about the relationship
between the {\hi} mass-to-light ratio and galaxy density over the
very large scales and very low overdensities we probe. For comparison,
galaxies in the Virgo Cluster, which has a stellar overdensity two
orders of magnitude higher than the field \citep{davies2012}, appear
to be four times more  {\hi}-deficient compared with background field
galaxies \citep{taylor12a}. Crudely assuming that the interaction
rate between galaxies (as for gaseous particles) is proportional to
the product of density and velocity dispersion, an overdensity of
30 per cent will only lead to an {\hi} deficiency of 1 per cent or
so.

\subsection{Binning\label{sub:Binning}}

\begin{figure*}
\includegraphics{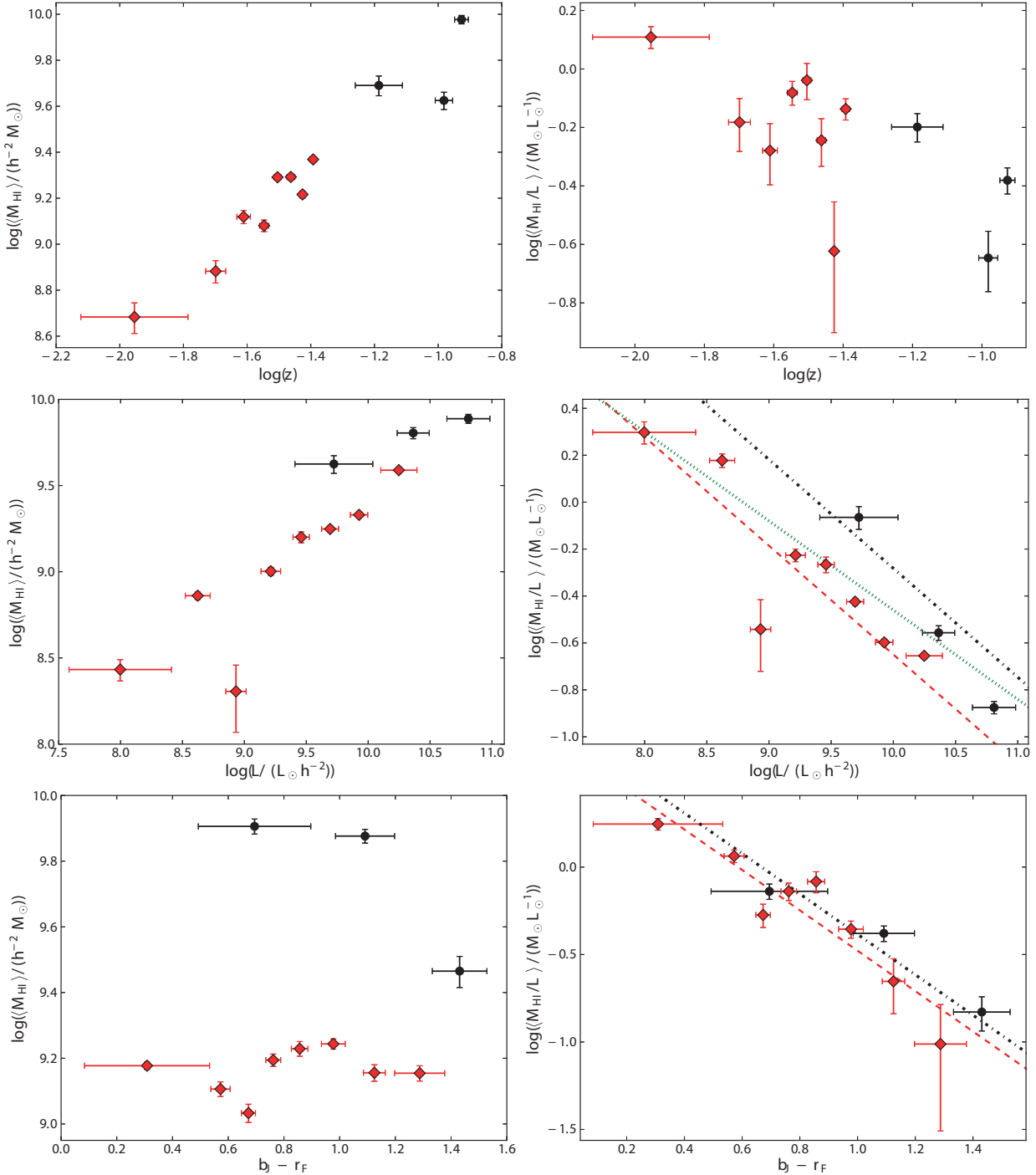}

\caption{The average {\hi} masses (left) and average mass-to-light ratios
(right) derived by co-adding all spectra in bins of (from top to bottom)
redshift, $b_{J}$ luminosity and $(b_{J}-r_{F})$ colour. The values
are plotted at the mean value of the binned parameter in each subsample.
The error bars represent the 1\,$\sigma$ spread within the bin.
Values derived from SGP data are shown as black points, while the
HIPASS values are shown as red diamonds. The red dashed lines in the
two lower-right panels show the least-squares fit to the HIPASS points.
The black dot-dashed lines show the fit to the SGP points, assuming
the same gradient as the HIPASS fit. The green dotted line shows the
relationship between $M_{{\rm \hi}}/L$ and luminosity as estimated
by \citet{karachentsev08}.}
\label{Flo:binned_plots}
\end{figure*}

\begin{table*}
\noindent \begin{centering}
\begin{tabular}{ccccccccc}
\hline 
\multicolumn{1}{|c|}{Property} & \multicolumn{1}{c|}{Bin range} & \multicolumn{1}{c|}{Mean} & \multicolumn{1}{c|}{rms} & \multicolumn{1}{c|}{$N$} & \multicolumn{1}{c|}{Integrated S/N} & \multicolumn{1}{c|}{Peak S/N} & \multicolumn{1}{c|}{$\left\langle M_{\hi}\right\rangle $$(\times10^{9}$\,$h^{-2})$} & \multicolumn{1}{c|}{$ $$\left<M_{\hi}/L'\right>$}\tabularnewline
\hline 
\hline 
$z$ & $0.0406-0.0863$ & 0.0651 & 0.0111 & 1082 & 10.2 & 4.8 & $4.90\pm0.48$ & $0.63\pm0.07$\tabularnewline
 & $0.0863-0.1116$$ $ & 0.1043 & 0.0066 & 1099 & 11.6 & 4.8 & $4.21\pm0.36$ & $0.22\pm0.05$\tabularnewline
 & $0.1116-0.1318$ & 0.1184 & 0.0059 & 1096 & 23.5 & 9.0 & $9.48\pm0.40$ & $0.42\pm0.04$\tabularnewline
\hline 
$\log$$\left(L'/({\rm L}{}_{\odot}\, h^{-2})\right)$ & $8.645-10.122$ & 9.723 & 0.314 & 1082 & 8.5 & 5.5 & $4.22\pm0.50$ & $0.86\pm0.10$\tabularnewline
 & $10.122-10.576$ & 10.364 & 0.129 & 1081 & 13.5 & 7.1 & $6.38\pm0.47$ & $0.28\pm0.02$\tabularnewline
 & $10.576-11.385$ & 10.810 & 0.174 & 1114 & 16.7 & 6.4 & $7.72\pm0.46$ & $0.13\pm0.01$\tabularnewline
\hline 
($b_{J}-r_{F}$)  & $-1.699-0.918$ & 0.695 & 0.2019 & 1083 & 19.0 & 10.6 & $8.05\pm0.42$ & $0.73\pm0.07$\tabularnewline
 & $0.918-1.271$ & 1.091 & 0.1061 & 1081 & 20.7 & 7.6 & $7.52\pm0.36$ & $0.42\pm0.04$\tabularnewline
 & $1.271-2.089$ & 1.431 & 0.0980 & 1113 & 9.2 & 3.0 & $2.92\pm0.32$ & $0.15\pm0.03$\tabularnewline
\hline 
\end{tabular} 
\par\end{centering}

\caption{Binned parameters for the SGP data. Columns (from left to right) are:
the binned property, bin range, mean and rms within the bin range,
number of sources, integrated and peak signal-to-noise ratio of the
stacked mass spectrum, average mass (units ${\rm M}_{\odot}$) and
average mass-to-light ratio (units ${\rm M}_{\odot}/{\rm L}_{\odot}$)
calculated from the stacked spectrum.}
\label{Tab:binning-SGP-1}
\end{table*}

\begin{table*}
\noindent \begin{centering}
\begin{tabular}{ccccccccc}
\hline 
\multicolumn{1}{|c|}{Property} & \multicolumn{1}{c|}{Bin range} & \multicolumn{1}{c|}{Mean} & \multicolumn{1}{c|}{rms} & \multicolumn{1}{c|}{N} & \multicolumn{1}{c|}{Integrated S/N} & \multicolumn{1}{c|}{Peak S/N} & \multicolumn{1}{c|}{$\left\langle M_{\hi}\right\rangle $$(\times10^{9}$\,$h^{-2})$} & \multicolumn{1}{c|}{$\left<M_{\hi}/L'\right>$}\tabularnewline
\hline 
\hline 
$z$ & $0.0026-0.0171$ & 0.0111 & 0.0043 & 1814 & 6.6 & 4.0 & $0.48\pm0.07$ & $1.28\pm0.11$\tabularnewline
 & $0.0171-0.0224$ & 0.0200 & 0.0014 & 1981 & 9.0 & 6.4 & $0.77\pm0.09$ & $0.66\pm0.13$\tabularnewline
 & $0.0224-0.0265$ & 0.0245 & 0.0012 & 1814 & 15.1 & 12.2 & $1.32\pm0.09$ & $0.53\pm0.12$\tabularnewline
 & $0.0265-0.0298$ & 0.0284 & 0.0009 & 1964 & 17.3 & 9.8 & $1.12\pm0.07$ & $0.83\pm0.08$\tabularnewline
 & $0.0298-0.0327$ & 0.0313 & 0.0008 & 1799 & 33.7 & 13.8 & $1.96\pm0.06$ & $0.94\pm0.13$\tabularnewline
 & $0.0327-0.0361$ & 0.0345 & 0.0010 & 1986 & 36.7 & 15.2 & $1.97\pm0.05$ & $0.57\pm0.11$\tabularnewline
 & $0.0361-0.0388$ & 0.0374 & 0.0008 & 1798 & 28.0 & 11.1 & $1.65\pm0.06$ & $0.24\pm0.11$\tabularnewline
 & $0.0388-0.0423$ & 0.0405 & 0.0010 & 1937 & 70.2 & 17.4 & $2.35\pm0.03$ & $0.73\pm0.06$\tabularnewline
\hline 
$\log$$\left(L'/({\rm L}{}_{\odot}\, h^{-2})\right)$ & $6.206-8.426$ & 7.997 & 0.413 & 1812 & 7.1 & 7.3 & $0.30\pm0.04$ & $1.98\pm0.201$\tabularnewline
 & $8.426-8.787$ & 8.623 & 0.101 & 1963 & 19.8 & 6.8 & $0.73\pm0.04$ & $1.51\pm0.10$\tabularnewline
 & $8.787-9.073$ & 8.932 & 0.082 & 1810 & 2.4 & 5.4 & $0.20\pm0.09$ & $0.29\pm0.10$\tabularnewline
 & $9.073-9.345$ & 9.214 & 0.079 & 1962 & 18.3 & 11.0 & $1.01\pm0.06$ & $0.59\pm0.04$\tabularnewline
 & $9.345-9.574$ & 9.459 & 0.066 & 1811 & 13.8 & 12.3 & $1.59\pm0.12$ & $0.54\pm0.04$\tabularnewline
 & $9.574-9.809$ & 9.683 & 0.069 & 1962 & 26.3 & 12.1 & $1.77\pm0.07$ & $0.37\pm0.01$\tabularnewline
 & $9.809-10.050$ & 9.927 & 0.070 & 1811 & 34.1 & 14.8 & $2.13\pm0.06$ & $0.25\pm0.01$\tabularnewline
 & $10.050-10.734$ & 10.248 & 0.146 & 1962 & 39.1 & 19.2 & $3.88\pm0.10$ & $0.22\pm0.01$\tabularnewline
\hline 
($b_{J}-r_{F}$)  & $-0.994-0.504$ & 0.309 & 0.224 & 1814 & 37.7 & 13.5 & $1.51\pm0.04$ & $1.76\pm0.13$\tabularnewline
 & $0.504-0.629$ & 0.572 & 0.035 & 1965 & 19.5 & 13.4 & $1.28\pm0.07$ & $1.16\pm0.10$\tabularnewline
 & $0.629-0.716$ & 0.673 & 0.025 & 1816 & 15.7 & 12.9 & $1.08\pm0.07$ & $0.53\pm0.08$\tabularnewline
 & $0.716-0.807$ & 0.762 & 0.027 & 1956 & 23.1 & 13.0 & $1.57\pm0.07$ & $0.73\pm0.08$\tabularnewline
 & $0.807-0.908$ & 0.856 & 0.030 & 1814 & 19.4 & 12.4 & $1.70\pm0.09$ & $0.83\pm0.11$\tabularnewline
 & $0.908-1.056$ & 0.977 & 0.043 & 1955 & 27.8 & 12.8 & $1.76\pm0.06$ & $0.44\pm0.05$\tabularnewline
 & $1.056-1.190$ & 1.125 & 0.039 & 1817 & 17.3 & 9.6 & $1.44\pm0.08$ & $0.22\pm0.08$\tabularnewline
 & $1.190-1.958$ & 1.287 & 0.090 & 1956 & 18.6 & 9.8 & $1.43\pm0.08$ & $0.10\pm0.07$\tabularnewline
\hline 
\end{tabular}
\par\end{centering}

\caption{Binned parameters for the HIPASS data. Columns are as defined in Table
\ref{Tab:binning-SGP-1}.}
\label{Tab:binning-hipass-1-1}
\end{table*}

\begin{table*}
\begin{tabular}{|c|c|c|c|c|c|}
\hline 
 & \multicolumn{2}{c|}{HIPASS} & SGP & \multicolumn{2}{c|}{\citep{karachentsev08}}\tabularnewline
$x$ & $a$ & $b$ & $b$ & $a$ & $b$\tabularnewline
\hline 
$\log$$\left(L'/({\rm L}{}_{\odot}\, h^{-2})\right)$ & $-0.46$ & 4.00 & 4.35 & $-0.38$ & $3.34$\tabularnewline
($b_{J}-r_{F}$)  & $-1.15$ & 0.68 & $0.77$ & - & -\tabularnewline
\hline 
\end{tabular}

\caption{Parameters for the lines in Fig. \ref{Flo:binned_plots} showing the
derived relationships between {\hi} mass-to-light ratio and luminosity
and colour: $ $$\log(\left<M_{\hi}/L'\right>)=ax+b$. The fit to
the SGP points has been derived using the HIPASS gradient; hence,
only the offset is quoted.\label{tab:fit_parameters}}
\end{table*}

We now investigate {\hi} trends with redshift, luminosity and colour.
We split the parent sample into bins containing roughly equal source
counts. We then stack the mass spectra of all sources within each
subsample. By splitting the SGP parent sample into three bins and
the HIPASS sample into eight bins, we can still achieve a statistically
significant detection in each. Tables \ref{Tab:binning-SGP-1} and
\ref{Tab:binning-hipass-1-1} show the properties of each SGP and
HIPASS subsample, respectively. The variation of the average {\hi}
mass and mass-to-light ratio in each bin is plotted in Fig. \ref{Flo:binned_plots}. 

To minimize confusion in our results, we have restricted the measurement
of both $\left\langle M_{\hi}\right\rangle $ and $\left\langle M_{\hi}/L\right\rangle $
to the $\pm300$\,km\,s$^{-1}$ spectral range, as in Section \ref{sub:Cosmic-HI-mass}.
We have adjusted the luminosities to account for confusion (via equation
\ref{eq:lum_corr}) before binning. We have also used these corrected
luminosities ($L'$) to determine the mass-to-light ratios. Note,
however, that it is not appropriate to apply the $M_{\hi}/L$ weighting
(equation \ref{eq:ML_weight}) to this binning analysis.

The average {\hi} mass-to-light ratios we measure are consistent
with the results of \citet{doyle05} who study HIPASS galaxies with
optical counterparts identified in the 6dF Galaxy Survey. They find
that the $b_{J}$-band mass-to-light ratios of these galaxies are
predominantly less than 5\,${\rm M}_{\odot}/{\rm L}_{\odot}$. For
comparison, the highest mass-to-light ratio we measure is $2.0\pm0.2$\,${\rm M}_{\odot}/{\rm L}_{\odot}$
and is associated with the lowest HIPASS luminosity bin.

We find that the {\hi} mass increases and the mass-to-light ratio
decreases with $b_{J}$ luminosity in both the HIPASS and SGP samples.
The former reflects the fact that more luminous galaxies are often
larger in size and therefore contain greater total {\hi} masses (eg.
\citealt{toribio11}). The fact that lower optical luminosity galaxies
have higher mass-to-light ratios has been well established through
direct {\hi} observations (eg. \citealt{warren06}) and we have verified
here that such trends can be reproduced with a stacking analysis. 

We have performed a least-squares linear fit to the HIPASS data, as
shown in Fig. \ref{Flo:binned_plots}. Fitting to the SGP data is
less robust, owing to the small number of bins. However, we perform
the linear fit assuming the same gradient as the HIPASS fit. The dotted
line in Fig. \ref{Flo:binned_plots} shows the close match of the
mass-to-light versus luminosity derived by \citet{karachentsev08},
compared with our results (see also Table \ref{tab:fit_parameters}).
This verifies our adjustment for the luminosity dependence of the
mass-to-light ratio. The only SGP point to substantially deviate is
the lowest luminosity bin, which is the main reason for the slightly
increased value of $\Omega_{\hi}$ in Table \ref{tab:OHI values}.

Similar {\hi} trends are seen with redshift. This does not reveal
any real evolution, but arises because we do not have a volume-limited
sample and are therefore preferentially detecting the higher mass
and luminosity galaxies at higher redshifts. It is precisely this
bias that we are attempting to correct via equation \prettyref{eq:ML_weight}
in our above calculation of the {\hi} mass density.

We use the SuperCOSMOS $b_{J}$ and $r_{F}$ magnitudes, as defined
in the 2dFGRS catalogue, to examine trends with colour. For the SGP
population, which contains high-mass galaxies in each colour bin,
a higher total {\hi} mass is seen for bluer sources. The HIPASS sample
contains a range of masses at all colours, and it is therefore unfair
to directly compare the {\hi} masses in each bin. The large vertical
offset between the two samples can again be explained by the preferential
inclusion of higher mass sources in the SGP sample due to the magnitude
limits of the optical catalogue. 

Taking the ratio with the optical luminosity removes the galaxy size
dependence, revealing a clear relationship between the {\hi} mass-to-light
ratio and colour. We see that bluer sources have higher mass-to-light
ratios. Again, we perform a linear fit to the HIPASS data, and apply
this gradient to the SGP fit. Both data sets are remarkably consistent
and reflect similar trends found between colour and gas fraction in
previous studies using both direct {\hi} detections (eg. \citealt{spitzak98})
and 21\,cm stacking experiments (\citealt{fabello11}). This supports
the notion that galaxies with larger gas supplies are generally associated
with increased star formation. 

Much of the colour dependence reflects the colour-magnitude relation
and is therefore corrected for in our calculation of $\Omega_{\hi}$,
which incorporates a luminosity correction to $M_{\hi}/L$. However,
it is possible that there remain residual differences between galaxies
in the SGP and HIPASS samples. For example, \citet{li12} examined
the residual dependence of {\hi}-to-stellar mass ratio on a number
of parameters, and found that a simultaneous dependence on stellar
mass and colour (as well surface brightness and colour gradient) best
explains the properties of luminous SDSS galaxies. However, after
application of the luminosity correlation in Table \ref{tab:fit_parameters},
we find no residual dependence of $M_{\hi}/L$ on colour to an accuracy
of 10 per cent over the mean colour difference between SGP and HIPASS
galaxies.

\section{Conclusions\label{sec:Conclusions-and-future}}

We have presented an {\hi} spectral stacking analysis of galaxies
identified in the 2dFGRS. We have shown that {\hi} stacking can be
used to efficiently and accurately probe the statistical properties
of high-redshift field galaxies.

Our sample consists of 15,093 galaxies at $0.0025<z<0.0423$ and 3,277
galaxies at $0.0405<z<0.1319$. 21\,cm data for the low-redshift
sample were provided by HIPASS. For the high-redshift sample, new
21\,cm observations of a 42\,deg$^{2}$ field near the SGP were
conducted with the Parkes radio telescope. Many of the low-redshift
optical galaxies are found to have strong {\hi} signatures, consistent
with the HICAT. Only one galaxy is detected above the 5\,$\sigma$
limit in the SGP sample.

We have co-added the {\hi} spectra of all galaxies in our sample,
after aligning each at the rest frame. We thus report a strong 31\,$\sigma$
average detection of the HIPASS galaxies and a 12\,$\sigma$ detection
for the SGP galaxies. The sensitivity level we achieve for our stacked
spectra is equivalent to observing a single object for $\sim37$\,d.
Therefore, we have shown that spectral stacking is an effective method
of making statistical detections of large numbers of individually
undetected galaxies at high redshift with reasonable integration times.

The rms noise in the stacked signal displays Gaussian behaviour, decreasing
with the square root of the number of co-added spectra. We find no
apparent intrinsic limitation in the depth to which we can continue
stacking experiments with the Parkes telescope. This bodes well for
future stacking experiments with deep, large-scale {\hi} surveys
on the next generation of radio telescopes such as the Australian
SKA Pathfinder, MeerKAT, APERTIF and ultimately the SKA.

We measure an average {\hi} mass of $\left\langle M_{{\rm \hi}}\right\rangle =(1.48\pm0.03)\times10^{9}$\,$h^{-2}\,{\rm M}_{\odot}$
from the HIPASS stacked spectrum and $\left\langle M_{{\rm \hi}}\right\rangle =(6.93\pm0.17)\times10^{9}$\,$h^{-2}\,{\rm M}{}_{\odot}$
from the SGP stacked spectrum. However, we find that these averages,
particularly that of the SGP sample, are over estimated due to source
confusion. High-resolution follow-up observations to combat this issue
are underway with the Australia Telescope Compact Array and will be
presented in a future paper. 

For now, we employ the Tully-Fisher relation to estimate source line
widths and thus to predict the extent of confusion in our data. We
then adjust the optical luminosities to include a beam-weighted sum
of all confused galaxies. Finding the average mass-to-light ratio
and accounting for sample bias, we derive the cosmic {\hi} mass density
($\Omega_{{\rm \hi}}$). We find $\Omega_{\hi}=(2.82{}_{-0.59}^{+0.30})\times10^{-4}\, h^{-1}$
at $0<z<0.04$ and $(3.19_{-0.59}^{+0.43})\times10^{-4}\, h^{-1}$
at $0.04<z<0.13$, i.e. we find no (12$\pm$23 per cent) evolution
over the past $\sim1$\,$h^{-1}$\,Gyr. Within the error margins,
these values agree with previous $\Omega_{{\rm \hi}}$ estimates made
via direct detections in the HIPASS, ALFALFA and AUDS surveys. We
argue that our results are far more robust to cosmic variance.

Finally, we employ the stacking technique to investigate the variation
of {\hi} properties with redshift, luminosity and colour. Trends
seen with redshift are obscured by selection effects. We find that
lower luminosity galaxies have lower total {\hi} masses but higher
mass-to-light ratios. We also see a decrease in the mass-to-light
ratio from blue to red galaxies. These results agree with direct observations.

\section*{Acknowledgements}

We are grateful to staff at the Parkes radio telescope for technical
support, particularly Ettore Carretti and Stacy Mader. Thanks to Morag
Scrimgeour and Chris Harris for observing assistance. The Parkes radio
telescope is part of the Australia Telescope National Facility which
is funded by the Commonwealth of Australia for operation as a National
Facility managed by CSIRO. We acknowledge the efforts of the 2dFGRS
and HIPASS collaborations. JD wishes to thank the International Centre
for Radio Astronomy Research for funding support, Laura Hoppmann and
Stefan Westerlund for coding assistance and Tobias Westmeier, Aaron
Robotham, Chris Power, Martin Zwaan, Marc Verheijen and Jason Prochaska
for useful discussions. Parts of this research were conducted by the
Australian Research Council Centre of Excellence for All-sky Astrophysics
(CAASTRO), through project number CE110001020. We acknowledge the
use of the \textsc{topcat} software in our data analysis (http://www.starlink.ac.uk/topcat/).

\label{lastpage}

\bibliographystyle{/Users/jacinta/Documents/PhD/mnras_template/mn2e}

\end{document}